# Superconductor-insulator transitions in three-dimensional indium-oxide at high pressures

Bar Hen [1*], Victor Shelukhin [1], Eran Greenberg [2,a], Gregory Kh. Rozenberg [1], Aharon Kapitulnik [3,4,5], Alexander Palevski [1]

[1] *Raymond and Beverly Sackler School of Physics and Astronomy, Tel-Aviv University, Tel Aviv 69978, Israel*
[2] *Center for Advanced Radiation Sources, University of Chicago, Chicago, Illinois 60637, USA*
[3] *Geballe Laboratory for Advanced Materials, Stanford University, Stanford, CA 94305, USA*
[4] *Department of Applied Physics, Stanford University, Stanford, CA 94305, USA*
[5] *Department of Physics, Stanford University, Stanford, CA 94305, USA*

Experiments investigating magnetic-field-tuned superconductor-insulator transition (HSIT) mostly focus on two-dimensional material systems where the transition and its proximate ground-state phases, often exhibit features that are seemingly at odds with the expected behavior. Here we present a complementary study of a three-dimensional pressure-packed amorphous indium-oxide (InOx) powder where granularity controls the HSIT. Above a low threshold pressure of ~0.2 GPa, vestiges of superconductivity are detected, although neither a true superconducting transition nor insulating behavior are observed. Instead, a saturation at very high resistivity at low pressure is followed by saturation at very low resistivity at higher pressure. We identify both as different manifestations of anomalous metallic phases dominated by superconducting fluctuations. By analogy with previous identification of the low resistance saturation as a "failed superconductor", our data suggests that the very high resistance saturation is a manifestation of a "failed insulator". Above a threshold pressure of ~6 GPa, the sample becomes fully packed, and superconductivity is robust, with $T_C$ tunable with pressure. A quantum critical point at $P_C$~25 GPa marks the complete suppression of superconductivity. For a finite pressure below $P_C$, a magnetic field is shown to induce a HSIT from a true zero-resistance superconducting state to a weakly insulating behavior. Determining the critical field, $H_C$, we show that similar to the 2D behavior, the insulating-like state maintains a superconducting character, which is quenched at higher field, above which the magnetoresistance decreases to its fermionic normal state value.

## Introduction

In spite of extensive experimental and theoretical studies, the subject of superconductor to insulator quantum phase transitions (SIT) [1-3] remains controversial and new experimental findings in this subject are still in great demand. One of the most studied systems is a thin film of amorphous indium-oxide (InOx) for which at a certain value of oxygen concentration, the transition from superconducting to insulating state occurs [4, 5]. Most of the studies of this system were performed on two-dimensional (2D) thin films where the tuning towards the SIT is obtained by systematic annealing. Typically, the critical behavior of physical parameters such as $T_C$ (critical temperature of superconductivity) and $R_C$ (sheet resistance) in the vicinity of the critical point, is studied using some continuous tuning parameter. Since the SIT is observed at low temperatures, the tuning parameter cannot be changed using annealing in-situ and, therefore, preparation and low temperature measurements of different samples is required.

For this reason, the most reliable experimental method for studying critical phenomena is achieved

[a] *Present address: Applied Physics Division, Soreq Nuclear Research Center (SNRC), Yavne 81800, Israel*

by bringing the system close to the critical point in a discrete manner using annealing, and then applying a continuous control parameter such as magnetic field at low temperatures to drive the system from the superconducting phase to the insulating phase continuously. The transition observed using this method is called magnetic-field-tuned SIT (HSIT) [5-9].

Magnetic-field-tuned phase transitions in 2D films of InOx were reported in numerous studies [5, 6, 9-11]. These studies reveal several common features, such as well-pronounced peaks in the magnetoresistance (MR) of the insulating phase, the existence of a critical magnetic field ($H_C$), which is manifested in a single crossing point of the MR isotherms, and observation of scaling behavior which is related to universal critical exponents. Recently, the theoretically predicted "failed superconductor" state [12] was reported in experimental studies of InOx and In-InOx 2D systems by observation of low temperature saturation of sub-Drude values of resistivity [13,11]. Although theoretically the saturation of resistivity at low temperatures was also predicted in 3D [12] there was no experimental evidence for that.

Studies of 3D InOx [4] revealed the existence of SIT. However, no detailed analysis of the observed quantum phase transition was performed. Since the thickness of the films was at most 2000 Å, the authors could not approach the quantum critical point of the disordered driven phase transition while remaining in 3D since the hopping or localization lengths exceeded the thickness of the samples. Moreover, no magnetic-field-tuned phase transition was reported in 3D InOx as well as no saturation of resistivity at low temperatures.

In this paper we provide a thorough investigation of SIT in 3D InOx employing pressure as a different method for tuning the transition. The application of pressure using diamond anvil cells (DACs) in our experiments served two objectives. The first, to tune the transition without the necessity of annealing, and the second, to enable measurements of much thicker InOx samples up to tens of microns which, for all practical purposes, are 3D.

Our 3D sample in proximity to the quantum critical point exhibits features which resemble those of a 2D SIT, one of them being a characteristic peak shaped magnetoresistance in the insulating side of the HSIT. In contrast to the 2D studies, the peaks are less pronounced and a single crossing point of the isotherms in the magnetoresistance is absent.

In addition, the saturation of resistivity at low temperatures, which was reported in 2D systems of InOx at the "failed superconductor" state, is observed in our 3D sample. The saturated resistivity values in our sample are larger than the value of the normal state resistivity indicating a "failed insulator" state.

**Experimental**

Micron thick InOx continuous layers of amorphous InOx were deposited onto glass substrates using an $In_2O_3$ pallet, in an oxygen atmosphere at a pressure of $5 \times 10^{-5}$ Torr and at a slow rate of ~0.5-1 Å/sec. Then, a "powder" of InOx was prepared by the mechanical removal of the film in the form of flakes from the substrates.

Pressure was exerted using miniature diamond anvil cells (DACs) [14] with diamond anvil culets of 300 μm. For our transport measurements, a pre-indented rhenium gasket was drilled and then filled and covered with a powder layer of 75% $Al_2O_3$ and 25% NaCl for electrical insulation. The sample of InOx was placed on top of the electrical insulation. A Pt foil with a thickness of 5-7 μm was cut into six triangular probes connecting between the sample and copper leads allowing the electrical transport measurements at elevated pressures. A few ruby fragments for pressure determination were located in the region between the Pt electrode tips overlapping the sample. No pressure-transmitting medium was used, but pressure is effectively transmitted to the sample upon compression by way of the surrounding insulating layer.

Synchrotron XRD measurements, at both compression and decompression cycles, were performed at room temperature up to 12.3 GPa at the beamline 13ID-D of APS (Argonne, IL, USA), with a wavelength of λ = 0.3344 Å, in angle-dispersive mode with patterns collected using a MAR-165 CCD detector. A LaB6 standard was used to determine the sample-to-detector distance, beam center, and detector tilts. The x-ray was approximately 4x3 microns in size. Liquid Nitrogen was used as the pressure-transmitting medium in these experiments. Ruby was used as a pressure gauge in these measurements as well. The image data were integrated using DIOPTAS [15]. XRD data at ambient temperature and pressure were collected from a powder which was placed between 2 glass slides.

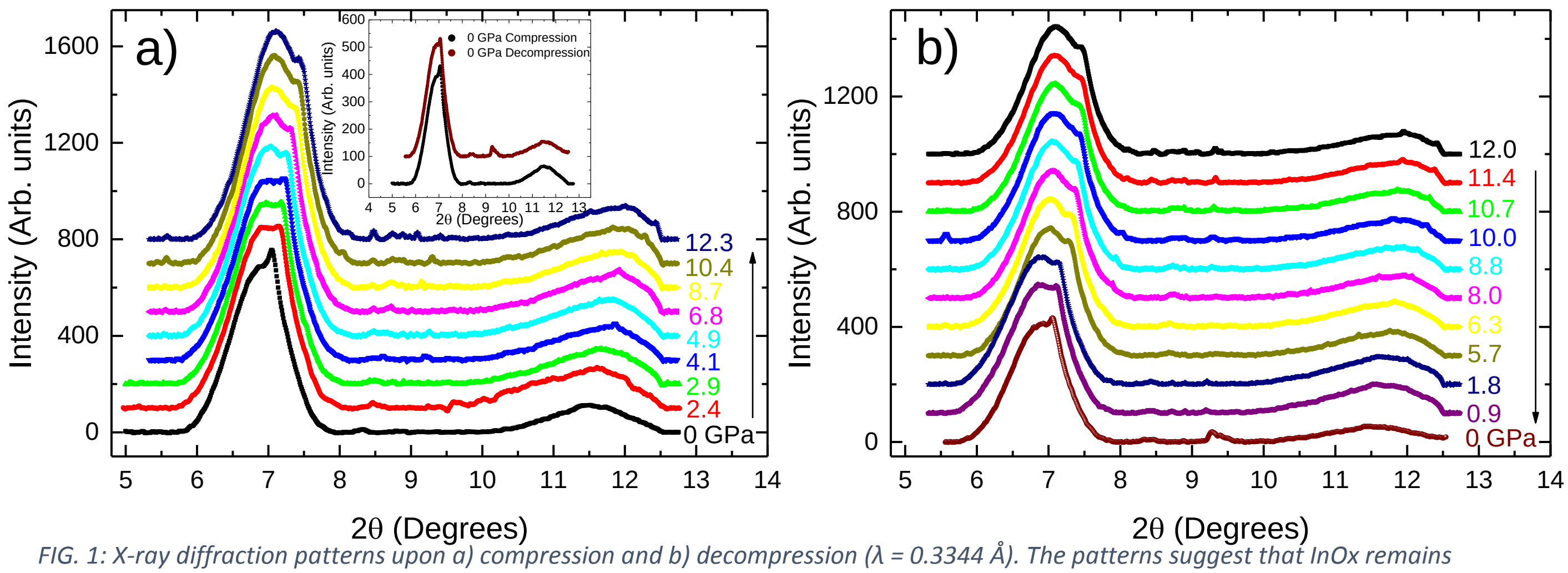

*FIG. 1: X-ray diffraction patterns upon a) compression and b) decompression (λ = 0.3344 Å). The patterns suggest that InOx remains amorphous at least up to 12.3 GPa. The inset shows that the pressure induced shift is reversible.*

Electrical transport measurements were performed in a dilution refrigerator. The sample was compressed up to 7.4 GPa, decompressed to 1.7 GPa and recompressed to 19.6 GPa at ~2 GPa intervals. After each pressure change, the sample cooled down from ambient temperature down to 0.04 K. At some of the pressure points, magnetic fields (up to 8 T) were applied at low temperatures.

**Experimental results**

In order to examine the structural changes of the sample at different pressures during the compression of InOx and the structure reversibility during the decompression, we performed XRD analysis at room temperature in a wide range of pressures up to 12.3 GPa. The diffraction patterns shown in Fig. 1(a) during compression suggests that InOx remains amorphous at least up to 12.3 GPa. The amorphous halo shifts gradually to larger angles, indicating a densification of the sample under pressure. The pressure-induced shift is reversible as can be seen in the decompression spectra in Fig. 1(b) and in the inset. The latter observation of the reproducibility is important and we will return to this point, when we discuss the observed irreversibility of the transport data for the first compression and decompression at low pressure range. Our transport measurements were performed in the range up to 19.6 GPa at compression, decompression and recompression. The details of the pressure ranges and the directions of pressure variations are provided below.

The resistance upon cooling down the sample during the first compression cycle are presented in Fig. 2. At the lowest pressure of 0.2 GPa, the sample is insulating since it exhibits a strong increase of the sample resistivity upon cooling, reaching unmeasurably high resistance values at very low temperatures. At a pressure of 1.6 GPa the sample shows signs of superconductivity as the resistance starts to drop at temperatures around 1 K. However, the resistance does not vanish at very low temperatures, but rather starts to increase upon continuation of cooling to lower temperatures and eventually saturates at ultra-low temperatures attaining a value which is higher than the normal state resistance. Further increase in pressure results in a further enhancement of superconductivity and at 4.8 GPa, the resistance saturates at sub-Drude values (Fig. 2 inset). For pressures exceeding 4.8 GPa the values of resistivity reach zero for vanishingly low temperatures. For pressures starting from 3.1 GPa we define $T_C$ based on half normal resistance value. The $T_C$ increases rather sharply from 0.2 K to 1.25 K with applied pressure in the range between 3.1 GPa and 5.8 GPa and then decreases to 1.2 K at 7.4 GPa as depicted in Fig. 3.

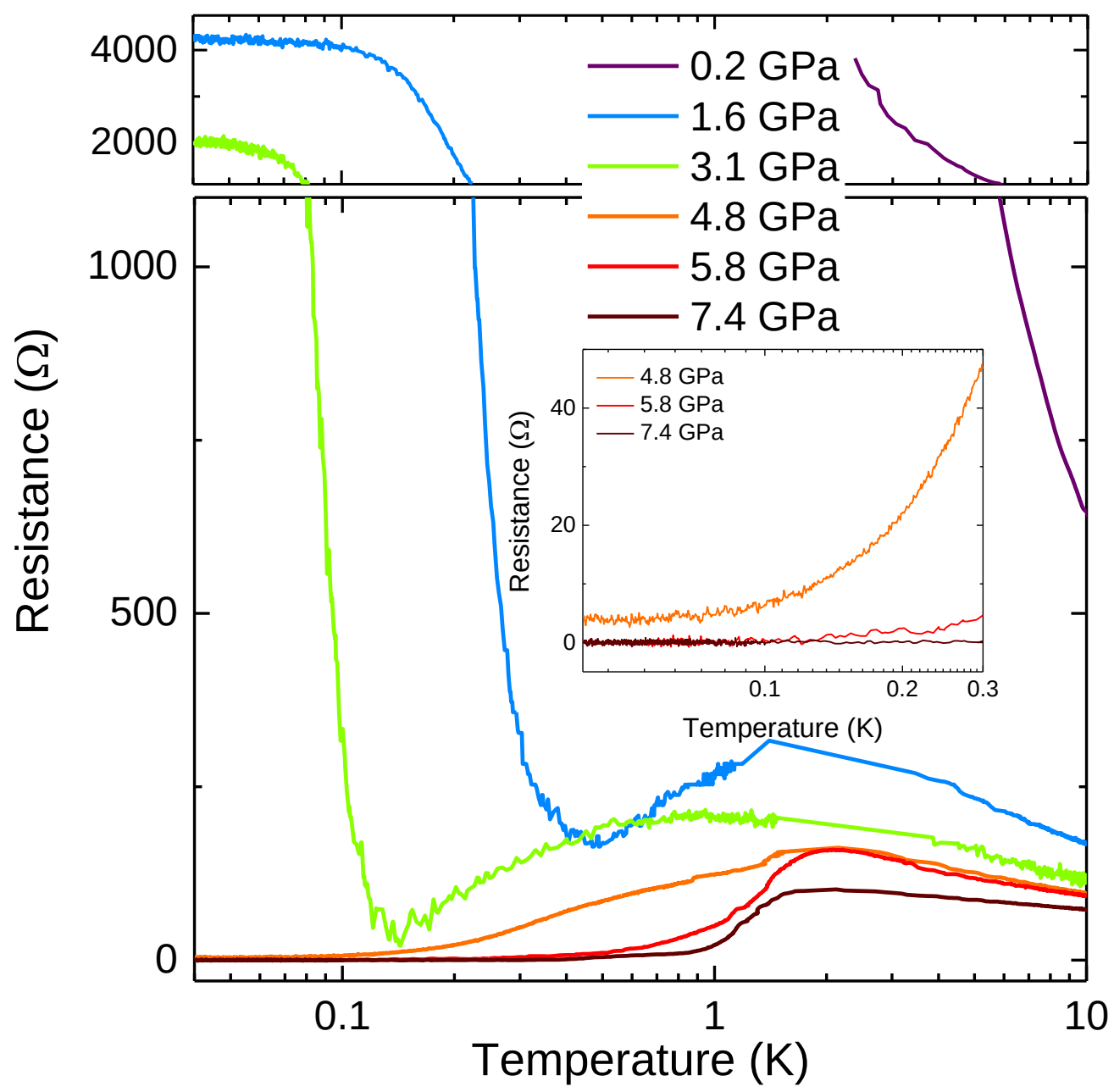

*FIG. 2: Temperature dependence of resistance at various pressures upon compression for the temperature range of 10 K down to 0.04 K. The graph shows Insulator to Superconductor transition and saturation at 1.6 GPa, 3.1 GPa and 4.8 GPa (inset).*

After compressing the sample to 7.4 GPa, we decompressed the sample down to 1.7 GPa and then recompressed it up to 19.6 GPa. We measured the cool-down curves for various pressures during these processes and the $T_C$ versus pressures is summarized in Fig. 3. The discrepancy between the $T_C$ values obtained from the first compression up to 5.8 GPa, and the values obtained from decompression and recompression clearly indicates irreversible changes occurring in the sample during the initial compression. We would like to remind the reader that there are no signs of any irreversibility noted in our XRD studies. After the initial compression, the $T_C$ of the compressed sample seems to follow a reversible variation curve. The reversible part of the data in Fig. 3 reveals the monotonic decrease of $T_C$ from 1.8 K at a low pressure of 1.7 GPa, to 0.36 K at a high pressure of 19.6 GPa. Therefore, while we leave for further discussions how the superconductivity arises in our sample during the first compression, it is obvious that the superconductivity is suppressed, manifested by diminishing $T_C$ upon increase of external pressure. Curiously enough, $T_C$ for pure In versus pressure within the low pressure range [16] has similar variation and might have similar microscopic origin.

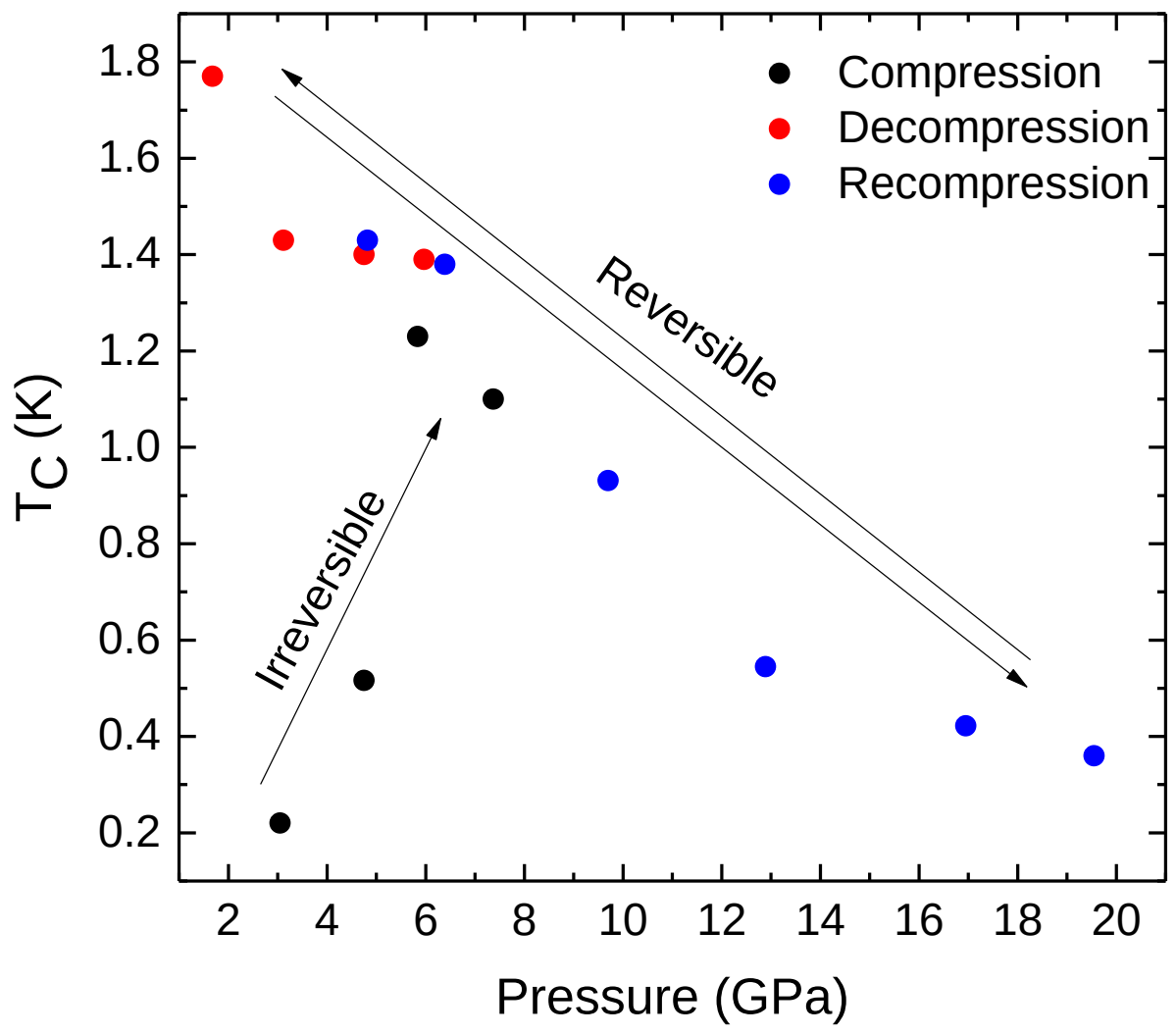

*FIG. 3: $T_C$ as a function of pressure upon compression (black), decompression (red) and recompression (blue). The diagram suggests that the initial compression to 5.8 GPa is irreversible.*

We now turn to our magnetoresistance measurements. Rather than presenting magnetoresistance data for all pressures, we focus on three specific pressures from different parts of the superconductivity versus pressure diagram of Fig. 3. The first, is at 1.6 GPa which is the first pressure at which the sample shows signs of superconductivity during the initial compression. Thus, this pressure is very close to the critical pressure ($P_C$) of the SIT from the "insulating" side. At this pressure, as can be seen in Fig. 4, the sample exhibits bell-shaped peaks. The peaks are broad and the resistance at the highest peak is larger by over 1 order of magnitude than the normal state resistance, namely, ~9 kΩ at 0.1 K at 6 T versus ~250 Ω above 1 K at zero magnetic field. Such peaks are usually associated with HSIT, which has been observed in thin films of InOx [5]. However, in 2D these peaks are much sharper. Nevertheless it is quite remarkable to see a similar shape of magnetoresistance curves in 3D.

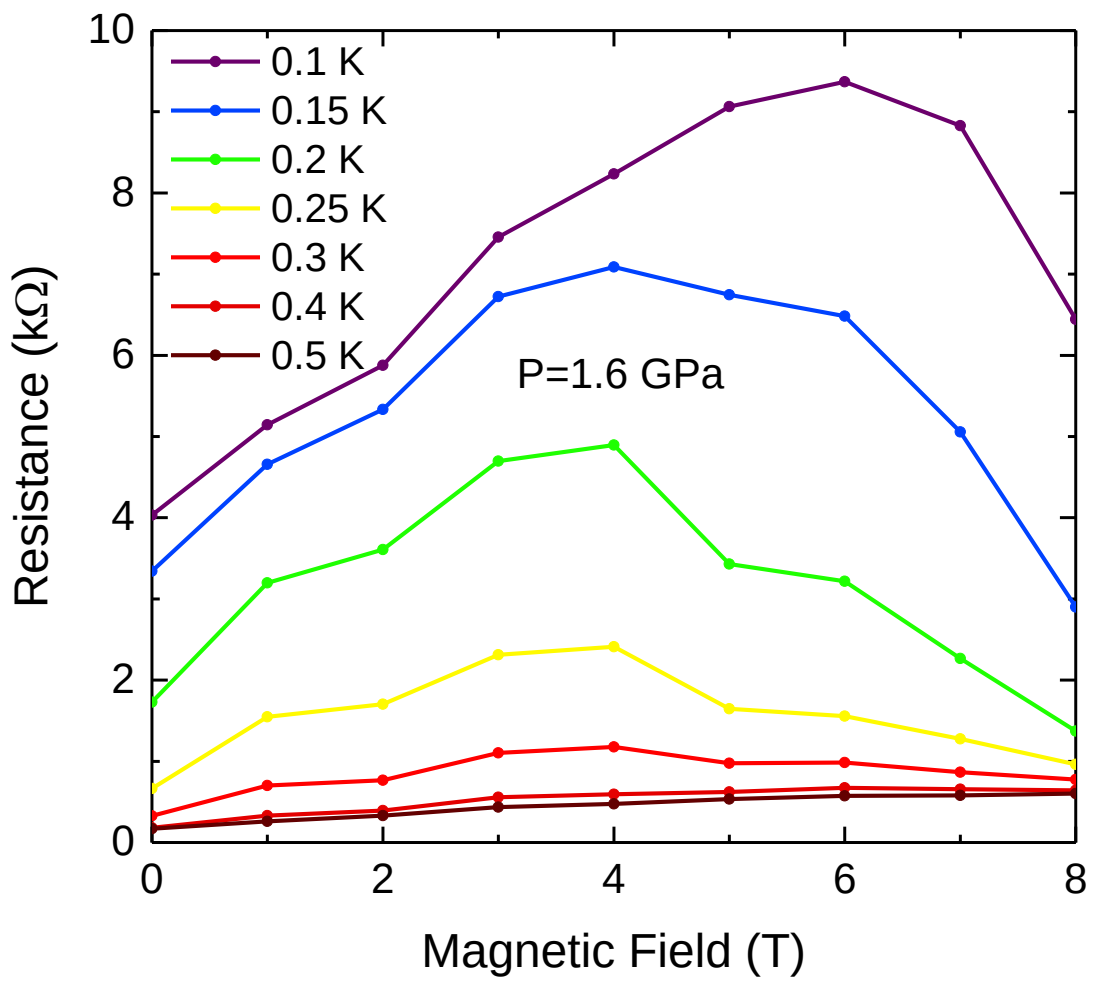

*FIG. 4: Peak shaped magnetoresistance isotherms at 1.6 GPa which are typical for HSIT.*

The second pressure that we present is 17.0 GPa (Fig. 5). At this pressure, the $T_C$ is quite low (0.42 K) and therefore the superconductivity at this pressure is quite suppressed, approaching the critical pressure for disappearance of the superconductivity from the superconducting side. Therefore, this pressure can be considered as being in proximity to another critical point at higher pressures. The graph shows magnetoresistance isotherms which are bell-shaped. Moreover, the cooling curves in the presence of magnetic fields exceeding 1 T clearly show insulating temperature dependence whereas the curves below 1 T show superconducting behavior as seen in the inset of Fig.5. This implies a HSIT in the vicinity of 1 T. The observation of a single crossing point of the magnetoresistance curves is a hallmark of 2D HSIT and is not expected for 3D systems. Indeed, instead of a single crossing point a rather broad region of different crossings of the isotherms is observed. We discuss this point further in the next section. The last set of magnetoresistance curves is presented for a pressure of 7.4 GPa in Fig. 6. This pressure point is sufficiently far from any SIT transition with quite high $T_C$ of 1.1 K.

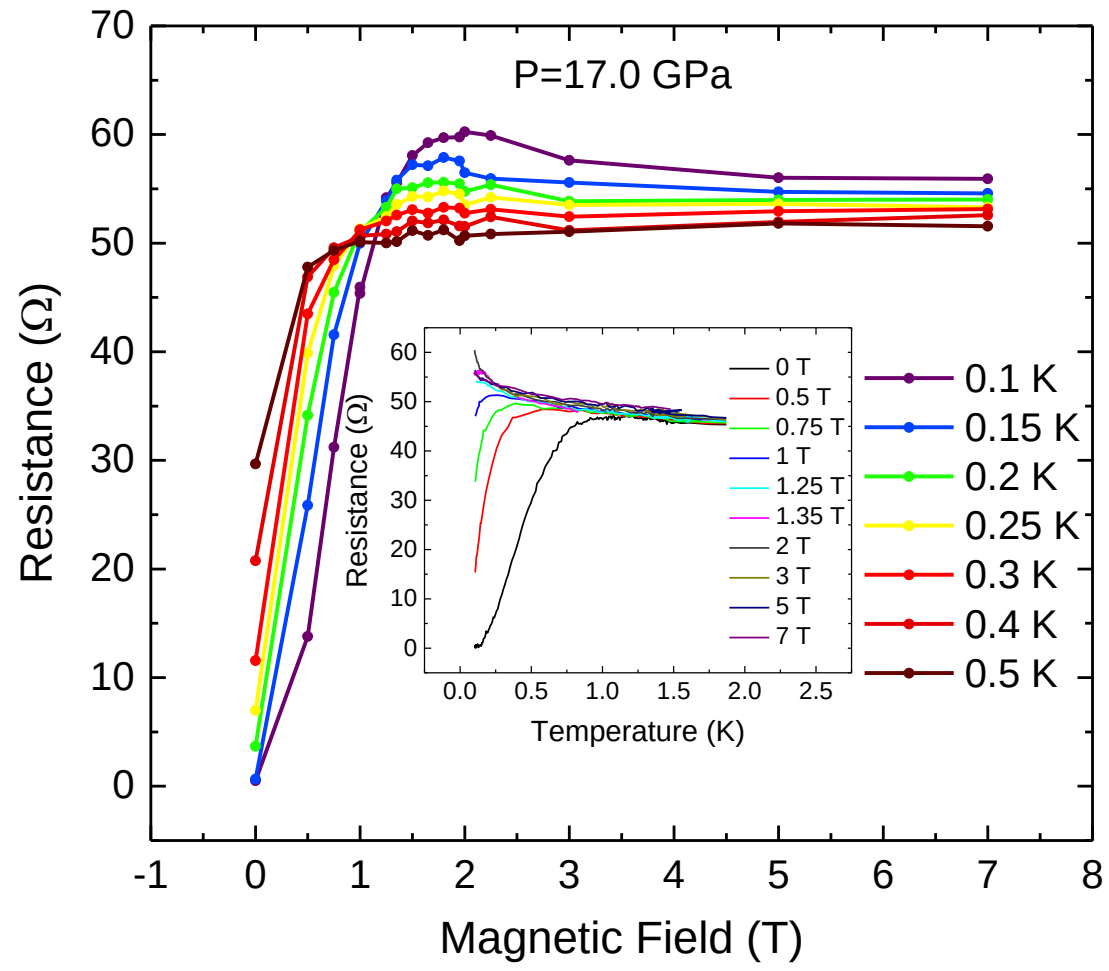

*FIG. 5:Magnetoresistance isotherms at 17.0 GPa. The curves at this pressure point are peak shaped, which is a hallmark of the HSIT. The crossing between each pair of isotherms occurs at different fields, signifying the absence of a single crossing point. The SIT is also shown in the inset where cool-down curves at various magnetic fields at this pressure are plotted.*

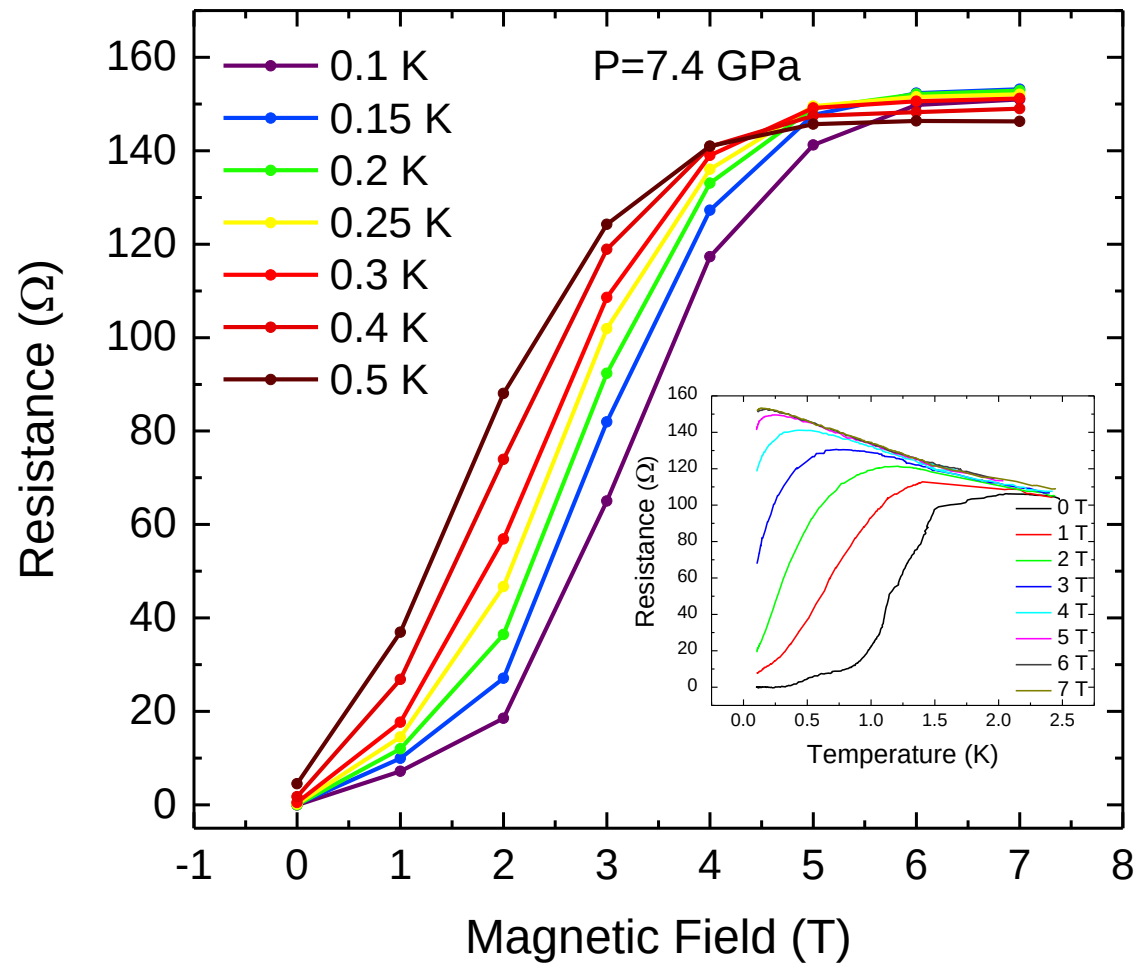

*FIG. 6: Magnetoresistance isotherms at 7.4 GPa. At this pressure point, which is sufficiently far from any transition, the magnetic field drives the sample from the superconductor to normal state. The inset shows the cool-down curves at various magnetic fields at this pressure.*

The magnetoresistance curves at this pressure, as can be seen in Fig. 6, exhibit a different behavior from the previous pressure points.

In contrast to the peaking shape, here the curves flatten. This implies that the magnetic field just drives the superconductor to a normal state.

Finally, we would like to focus on the data taken at the ultralow temperature range, down to 0.04 K, at 1.6 GPa. As we already pointed out above, at this pressure our InOx sample exhibits a decrease in the resistance in the vicinity of 1 K followed by an increase at lower temperatures as depicted in Fig. 2. In Fig. 7, we show the cooling curves at the same pressure of 1.6 GPa, taken for different magnetic fields. As clearly seen in Fig. 7, the rather sharp initial increase of the resistance upon cooling is followed by saturation behavior within the ultralow temperature range. The temperature below which the saturation starts, as well as the saturated value of the resistance depends on the applied magnetic field. The inset of Fig. 7 shows the linear variation of saturated resistance versus magnetic field.

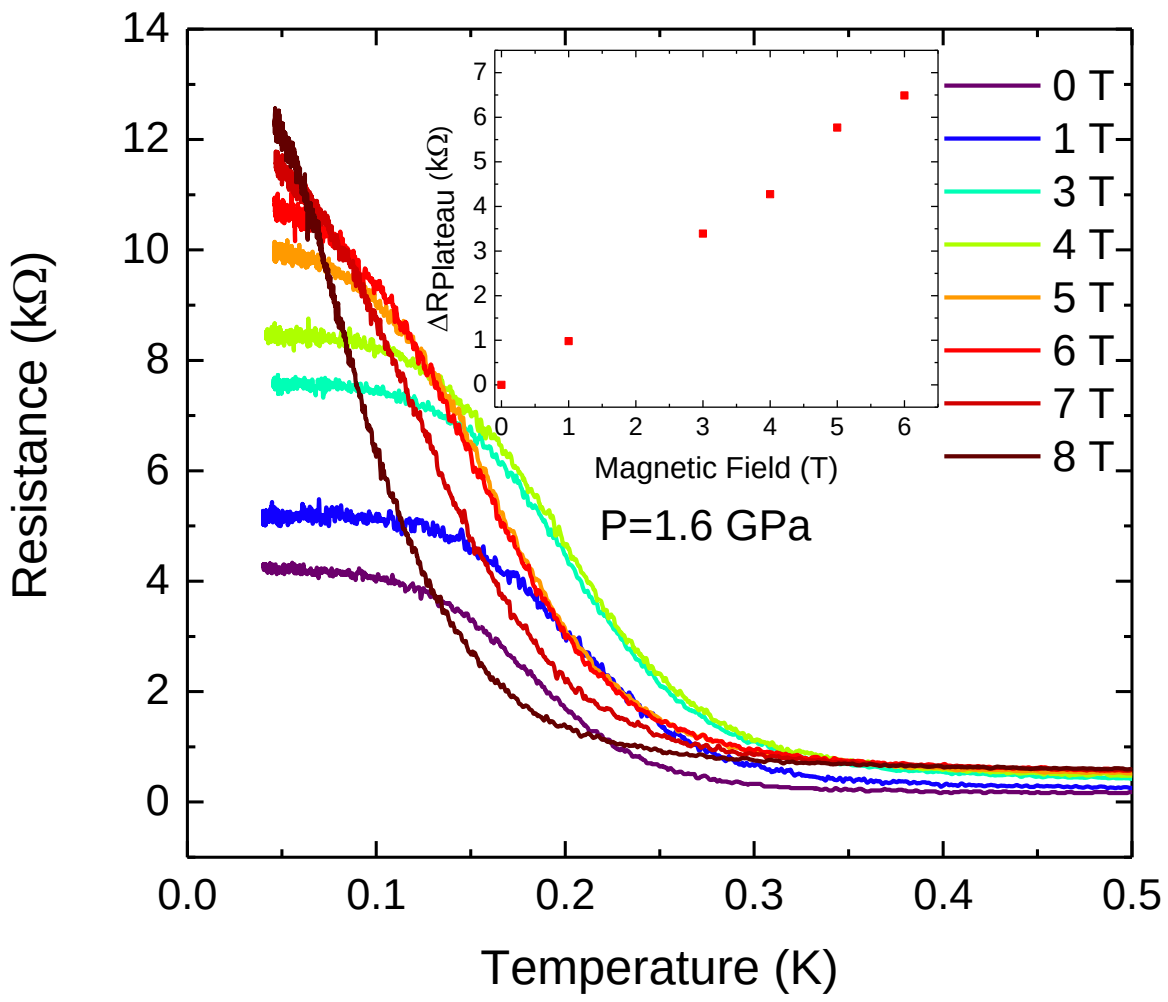

*FIG. 7: Temperature dependence of resistance at 1.6 GPa at various magnetic fields for the low temperature range of 0.5 K down to 0.04 K. The graph shows field dependent plateaus or saturations of resistance within the lowest temperature range. The inset shows the dependence in a more qualitative way.*

The temperature dependence of the resistance at 1.6 GPa for different magnetic fields in the range from 0 to 8T is presented in Fig. 8 on a semi logarithmic scale. The linear slopes of the curves in Fig. 8 suggest a hopping mechanism of the conductivity. Mott variable range hopping mechanism seems to fit the best the resistance variation for all magnetic fields versus temperature at a wide temperature range prior to saturation. The observed saturation and its significance will be discussed below in the next section of the paper.

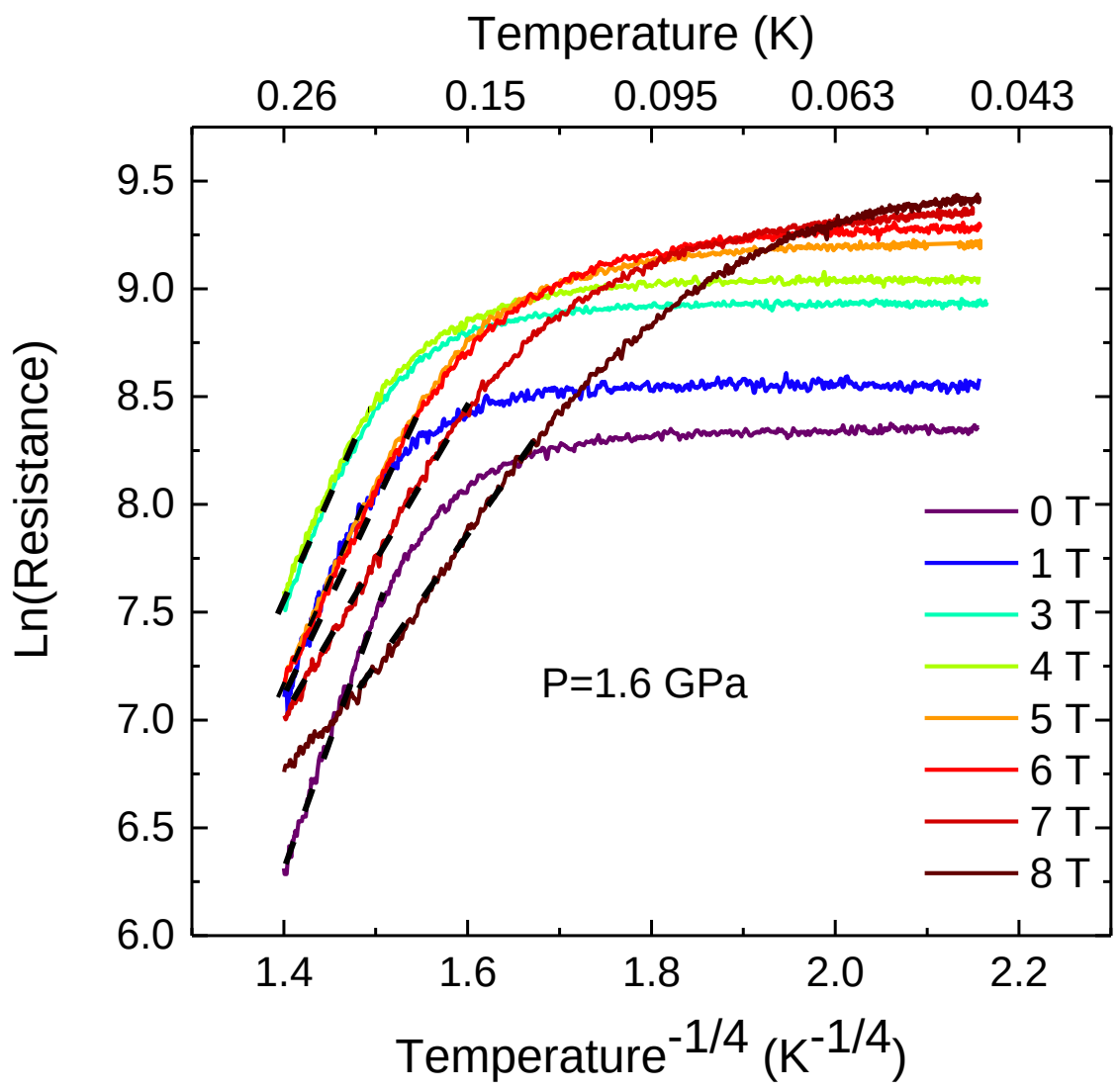

*FIG. 8: Resistance as a function of temperature to the power of $-\frac{1}{4}$ in a semi logarithmic scale. The linear slopes of the curves (indicated by dashed lines) imply that the conductivity follows the Mott variable range hopping mechanism.*

## Discussion and Analysis of the results

We would like to start the discussion of the presented data from the observed irreversibility of the initial compression exceeding 3.1 GPa which is clearly demonstrated in the transport measurements (Fig. 3) and is not observed in XRD data. Considering possible non reversible structural transformations taking place in the sample during the first compression cycle, we would like to note first a compacting of the sample due to elimination of voids between the flakes and shrinking of low-electron-density regions inside flakes (vacancy-like or void-like entities of various sizes, randomly distributed inside the sample (see, for example, [17]). Such irreversible sample compacting at the initial pressure application contributes to a better electrical coupling between the flakes influencing the localization of trapped electrons inside the flakes. We note also that the oxygen stoichiometry of the sample could be sensitive to the pressure application and may change during the initial compression. It is plausible that the oxygen concentration is higher when the sample consists of flakes with many voids in between them and inside, and therefore has a large surface relative to the compressed sample, where the surface to volume ratio is much smaller. Any combination of the above mentioned mechanisms would produce irreversibility in transport measurements which would not be detected by XRD. Regardless of the processes responsible for the irreversible part of the diagram, we believe that the SIT reported here belongs to the same universality class of quantum phase transitions with the pressure playing the role of the tuning parameter. This then adds to the list of other previously used tuning methods such as annealing gating or magnetic field, which may control any combination of carrier density, local superconducting gap and internal phase coherence.

We now turn back to the initial application of pressure, before the sample is fully "packed", thus exhibiting irreversible occurrence of superconductivity. This regime, which exists up to an initial packing pressure of ~5.8 GPa, is dominated by the inhomogeneous nature of the sample, which is expected to consist of a wide distribution of regions with varying strength of superconductivity. As is evident from Fig. 2, samples cooled at 1.6 GPa and 3.1 GPa show an insulating behavior, followed by a weak feature associated with the superconducting grains, and then continue to increase in resistance before saturating at low temperatures. Similar behavior was previously observed in thin Pb and Ga films as a function of their average thickness as they were deposited on a substrate [18]. There, as well, the initial film is highly inhomogeneous, full of voids, thus conduction is dominated by a tenuous path of grains, leading to saturation at low temperatures. While an insulating behavior, which largely avoids the superconducting grains was expected, the saturation to a value much larger than the quantum of resistance of $\frac{h}{e^2} \approx 25.8$ kΩ was a surprise.

A similar phenomenon emerges in our sample in the "irreversible" pressure regime, except that our sample is three-dimensional. For example, at 1.6 GPa (Fig. 7) the resistance at zero magnetic field saturates at ~ 4 kΩ, which implies a resistivity of ~1 Ω-cm, much larger than ~0.03 Ω-cm, which is the expected metal to insulator transition for amorphous InOx [19]. For such high resistivities normal InOx strongly diverges with decreasing temperature. A reasonable assumption is that vestiges of the superconducting regions within the otherwise insulating matrix, observed as a weak resistance reduction at ~1.5 K, prevent the resistance from diverging at low temperatures. However, such a simplistic approach would still predict low temperature divergence as long as the superconducting regions do not overlap or Josephson couple. In an attempt to overcome this problem, a composite model was proposed [20], where a certain set of parallel and series superconducting grains are connected through either normal or Josephson couplings, while the bare resistance of each junction is randomly distributed [20]. Indeed, this model showed resistance saturation resembling that observed in Pb and Ga films of [18]. A more realistic approach may be constructed based on Ambegaokar et al.'s model for hopping conductivity in disordered systems [21], previously used to explain the variation of $T_C$ in granular materials [22]. We assume a set of isolated superconducting grains with similar $T_C$, which are connected through weak links that are chosen from a random distribution. When the links are strong, they will be dominated by the Josephson coupling, when they are weaker, the thermal energy may break them to become normal, and when mostly weak, many are practically insulators, as is e.g. realized in the not fully packed grains. At high packing pressure, when true zero resistance is observed, we may assume that the system is dominated by the Josephson couplings and thus is expected to exhibit a true magnetic-field tuned superconductor-insulator transition. However, with low packing pressure the connections among grains are tenuous, dominated by non-conducting barriers and intergrain charging energy. Recall that in the superconducting regime, the standard theory can only account for a superconductor to an insulator transition, and the appearance of an intervening anomalous metallic phase, which is described as a "failed superconductor", needed careful consideration [12]. By analogy, within the dual behavior we expect that in the insulating regime, charge fluctuations that result from the coulomb blockade [23] in the more insulating barriers, will affect the coupling of grains, preventing a true insulating phase to form. In that respect we may dub this high-resistance anomalous metallic phase a "failed insulator". Indeed, similar to our zero-field data of Fig. 7, such a behavior was previously observed in two-dimensional arrays of small Josephson junctions, where the charging energy $E_C$ of the junctions cannot be neglected compared to either the Josephson coupling $E_J$ [24]. Flattening of the resistance at zero magnetic field in the insulating ("normal") state was attributed to quantum fluctuations associated with energy uncertainty of states with definite charge [23], where saturation at higher resistance was obtained for a larger ratio of $\alpha = \frac{E_C}{E_J}$. By analogy we expect that for the sample packed at 3.1 GPa the Josephson intergrain coupling will be stronger than at 1.6 GPa thus exhibit a lower resistance saturation as clearly shown in Fig. 2. Further increase in the Josephson intergrain coupling first leads to saturation at sub-Drude values ("failed superconductor") followed by a transition to a fully superconducting state as can be seen in the inset of Fig. 2.

Turning to the magnetic field tuned effect, both, one-dimensional [25] or two-dimensional [26] arrays exhibit a field tuned transition. While saturation of the resistance occurs at very low fields on the superconducting side and at high fields on the insulating side, it is not clear whether either of this can be attributed to quantum fluctuations in the thermodynamic limit (suggested in [25]), or it is a finite size effect (suggested in [26]). However, it is rather clear that unlike the arrays, our sample is three dimensional, macroscopic, and very well thermally anchored within the pressure cell. Thus, the progression from the low pressure packed sample dominated by the charging energy to the high pressure packed sample dominated by Josephson energy can be viewed as a gradual progression of the ratio α. Application of magnetic field on the 1.6 GPa sample shows the progression of the insulating state, where the resistance increases in some exponential fashion. Assuming that the resistance saturation in this regime is dominated by charge fluctuations which disrupt the tendency for a true disorder-dominated insulator, the application of magnetic field induces vortices in the superconducting regions, which move and generate voltage that acts as a driving force that suppresses the charge fluctuations. Fitting the resistance increase to an activated behavior where $R = R_0 e^{\left(\frac{T}{T_0}\right)^{\delta}}$, we find that it fits well with δ in the range of $\frac{1}{2}$ to $\frac{1}{4}$, which would agree with either coulomb-dominated or single-energy dominated variable range hopping. Fig. 8 shows an example fit with $\delta = \frac{1}{4}$, which we used to demonstrate the progression of the insulating state.

Further increase of the packing pressure increases the Josephson couplings among superconducting regions until a true superconducting state is observed above ~6 GPa. After reaching high packing, superconductivity can be tuned with pressure in a reversible manner. While the highest $T_C$ is now obtained upon removing most of the pressure, thus indicating a well packed sample, increasing the pressure decreases $T_C$, which seems to ultimately result in a quantum critical point, $P_C$, at ~25 GP (extrapolated from Fig. 3) where superconductivity disappears. We conjecture that this transition is driven by a faster increase in charging energy over the Josephson coupling and will be a subject of further investigation as it needs higher pressure. At any intermediate pressure a magnetic-field tuned transition is observed. The example shown in Figs. 5 and 6 demonstrate the effect of intergrain couplings, where stronger magnetoresistance is observed at higher pressure (Fig. 5) over the weak magnetoresistance at lower pressure (Fig. 6). Unlike the two-dimension counterpart, where the crossing point defines the value of the $H_C$ of the HSIT, there is no crossing point in 3D. The reason for the absence of a crossing point in 3D is related to the temperature dependence of the critical resistivity $\rho_C(T)$ separating the superconducting and insulating phases of the sample. Therefore, the crossings between different pairs of isotherms occur at different magnetic fields. The magnetic field at which the separatrix, $\rho_C(T)$, is achieved defines $H_C$.

Our InOx at 17.0 GPa clearly exhibits HSIT as manifested in the inset of Fig. 5 by the existence of a critical magnetic field, separating between superconducting and insulating curves. Although the value of $H_C$ cannot be determined exactly, based on the extrapolation of the curves Fig. 5 to T→0 it is within the range between 1 T to 1.25 T.

## Conclusions

We performed low temperature studies of 3D InOx films at elevated pressures. We observed the transition of InOx from the insulating phase at 0.2 GPa to a superconducting phase at 4.8 GPa. This transition is observed during irreversible initial compression and it occurs via an intermediate "failed insulator" state with finite temperature independent (saturated) value of the resistivity at zero temperature. We demonstrated that the increase of the pressure reduces $T_C$ in a reversible manner from 1.8 K at 1.7 GPa to 0.36 at 19.6 GPa. Extrapolating the variation of $T_C$ at high pressures in Fig. 3 yields a $P_C$ of ~25 GPa where superconductivity should vanish. In the vicinity of the critical pressure, at 17.0 GPa (Fig. 5), a HSIT with $H_C$ in the range between 1 T to 1.25 T is observed. In comparison with the 2D case, the HSIT does not exhibit a crossing point and has less pronounced peaks in the magnetoresistance. In addition, we have observed that the value of saturated resistance in the "failed insulator" state in our 3D sample can be tuned by a magnetic field in a similar fashion to 2D InOx in the "failed superconductor" state.

**Acknowledgements:**

The authors would like to thank Boris Spivak and Steve Kivelson for very helpful discussions.

This research was partially funded by PAZI foundation under grant number 326-1/20. Support from the Israeli Science Foundation (grant numbers 1748/20, and 227/16), is gratefully acknowledged.

Portions of this work were performed at GeoSoilEnviroCARS (The University of Chicago, Sector 13), Advanced Photon Source (APS), Argonne National Laboratory. GeoSoilEnviroCARS is supported by the National Science Foundation – Earth Sciences (EAR – 1634415) and Department of Energy- GeoSciences (DE-FG02-94ER14466). This research used resources of the Advanced Photon Source, a U.S. Department of Energy (DOE) Office of Science User Facility operated for the DOE Office of Science by Argonne National Laboratory under Contract No. DE-AC02-06CH11357.

Work at Stanford University was supported by the National Science Foundation Grant NSF-DMR-1808385.